\documentclass{agujournal}
\journalname{Space Weather}

\usepackage{textcomp}
\usepackage[varg]{txfonts}
\usepackage{multirow}
\usepackage{hhline}
\begin{document}

\title{Modeling the evolution and propagation of the 2017 September 9th and 10th 
	CMEs and SEPs arriving at Mars constrained by remote-sensing and in-situ measurement}

 \authors{Jingnan Guo\affil{1}, %\thanks{},
 Mateja Dumbovi\'c\affil{2},
 Robert F. Wimmer-Schweingruber\affil{1},
 Manuela Temmer\affil{2}, 
 Henning Lohf\affil{1},
 Yuming Wang \affil{3},
 Astrid Veronig\affil{2},
 Donald M. Hassler\affil{4}, 
 Leila M. Mays\affil{5}, 
 Cary Zeitlin\affil{6}, 
 Bent Ehresmann\affil{4},
 Olivier Witasse\affil{7},
 Johan L. Freiherr von Forstner\affil{1},
 Bernd Heber\affil{1},
 Mats Holmstr\"om\affil{8},
 Arik Posner \affil{9}
}

\affiliation{1}{Institut fuer Experimentelle und Angewandte Physik, University of Kiel, Kiel, Germany.}
\affiliation{2}{Institute of Physics, University of Graz, Austria.}
\affiliation{3}{University of Science and Technology of China, Hefei, China.}
\affiliation{4}{Southwest Research Institute, Boulder, CO, USA.}
\affiliation{5}{NASA Goddard Space Flight Center, USA.}
\affiliation{6}{Leidos, Houston, Texas, USA.}
\affiliation{7}{European Space Agency, ESTEC - Science Support Office, The Netherlands.}
\affiliation{8}{Swedish Institute of Space Physics, Kiruna, Sweden.}
\affiliation{9}{NASA Headquarters, Science Mission Directorate, Washington DC, USA}

\correspondingauthor{J. Guo}{guo@physik.uni-kiel.de}
\correspondingauthor{M. Dumbovi\'c}{mateja.dumbovic@uni-graz.at}

\begin{keypoints}
% Each must be 100 characters or less with no special characters or punctuation 
%\item Interplanetary Coronal Mass Ejections (ICMEs) arriving Mars and the usage of Forbush decreases for identifying their arrivals
%\item Graduated Cylindrical Shell (GCS) modeling of CME 3D geometry
%\item Comparison of \textit{in-situ} measurement with predictions from ICME propagation models such as WSA-ENLIL and Drag Based Model

%\item The 2017-09-10 CMEs and shock reached Mars and were detected by MSL-RAD as a large Forbush decrease
%\item The CME and shock early 3D kinematics are modeled using coronagraph images from SOHO and STEREO-A
%\item These combined with the Drag Based Model are used to derive the CME interaction and arrival at Mars 

%\item The 2017-09-10 CMEs \& shock produced one of the largest Forbush decrease seen by MSL-RAD since 2012
%\item The CME and shock early 3D kinematics are modeled using coronagraph images from SOHO and STEREO-A
%\item Characteristics of the shock \& CME arrival at Mars are compared with those using a Drag Based Model

%\item The Sept 2017 CMEs \& shock caused one of the largest Forbush decreases ever seen by MSL-RAD on Mars
%\item The ICME is extremely complex and is important for studying extreme space weather conditions at Mars
%\item Observation-based modeling of the ICME from its solar origin to its arrival at Mars are essential

%\item The 10 Sept 2017 SEP event was the first GLE observed at both Earth and Mars.
\item The 2017-09-10 SEP event was the first GLE observed on the surface of two different planets: Earth and Mars.
\item The SEP and ICME impact on Mars is helping us better understand extreme space weather conditions at Mars.
\item Synergistic modeling of the ICME and SEP propagation advances our understanding of such complex events for improved space weather forcasts.
\end{keypoints}

\begin{abstract} 
On 2017-09-10, solar energetic particles (SEPs) originating from the active region 12673  \replaced{were registered as a ground level enhancement (GLE) at Earth and the biggest GLE on the surface of Mars as observed by the Radiation Assessment Detector (RAD) since the landing of the Curiosity rover in August 2012}{produced a ground level enhancement (GLE) at Earth. The GLE on the surface of Mars, $\sim$ 160$^\circ$ longitudinally east of Earth, observed by the Radiation Assessment Detector (RAD) was the largest since the landing of the Curiosity rover in August 2012.} 
Based on multi-point coronagraph images\added{ and the Graduated Cylindrical Shell (GCS) model}, we identify the initial 3D kinematics of an extremely fast \added{Coronal Mass Ejection} \replaced{CME}{(CME)} and its shock front\added{,} as well as another \replaced{2}{two} CMEs launched hours earlier with moderate speeds\deleted{ using the Graduated Cylindrical Shell (GCS) model}. 
\replaced{These}{The} three CMEs interacted as they propagated outwards into the heliosphere and merged into a complex interplanetary CME (ICME). 
The arrival of the shock and ICME at Mars caused a very significant Forbush Decrease (FD) seen by RAD only a few hours later than that at Earth\added{,} which \replaced{is}{was} about 0.5 AU closer to the Sun. 
We investigate the propagation of the three CMEs and the \replaced{consequent}{merged} ICME together with the shock\added{,} using the Drag Based Model (DBM) and the WSA-ENLIL plus cone model constrained by the in-situ \replaced{SEP and FD/shock onset timing}{observations}.  
The synergistic \replaced{modeling}{study} of the ICME and SEP arrivals at Earth and Mars suggests that \deleted{in order }to better predict {potentially hazardous} space weather impacts at Earth and other heliospheric locations {for human exploration missions}, it is essential to analyze 1) the CME kinematics, especially during their interactions and 2) the spatially and temporally varying heliospheric conditions, such as the evolution and propagation of the stream interaction regions. 

\end{abstract}

\section{The flare, CMEs and GLE 72: close to the Sun}\label{sec:Sun}
During the declining phase of solar cycle 24, \added{from the 6th to 10th of September 2018 }heliospheric activity suddenly and drastically increased when the complex Active Region (AR) 12673 \added{located at the western solar hemisphere, }produced \replaced{three}{four} X-class flares and several Earth-directed Coronal Mass Ejections (CMEs) \citep{redmon2018} \deleted{from the 6th to 10th of September when it was located at the western solar hemisphere.}
The X9.3 flare on 2017-09-06 at S09W34 started at 11:53 UT, impulsively reached its peak {in the GOES soft X-ray flux} at 12:02 UT. \replaced{and}{It} was registered as the \replaced{biggest}{largest} flare of solar cycle no. 24. %and then ended at 12:10 UT. It

\subsection{The September 10th flare, flux rope and initial acceleration of particles}\label{sec:flare}

On {2017-09-10}, the same AR produced an X8.2 flare {at S08W88} (being the second largest one {of Cycle 24}) starting around 15:35 UT and {peaking} at 16:06 UT %in the GOES soft X-ray flux 
\citep{seaton2018observations, li2018spectroscopic, warren2018spectroscopic, long2018plasma,jiang2018decipher}. 
The flare was located on and slightly behind the west limb of the Sun as seen from Earth.
{As shown in Fig. \ref{fig:theSun}(a)-(b)}, remote sensing observations of the solar corona show that a magnetic flux rope (MFR) associated with the energetic flare started emerging at about 15:50, rose rapidly and triggered a fast eruption {starting from} about 15:53. 
{It was later observed as a CME in the white light (WL) coronagraph images of both the Solar Terrestrial relations observatory Ahead \citep[STA,][]{howard2008} and the Solar and Heliospheric Observatory \citep[SOHO,][]{brueckner1995large} as shown in Fig. \ref{fig:theSun}(e).} %(though more towards the {south of the ecliptic plane})
{References and descriptions of all the measurements and databases employed in this study are given in the Appendix.}

Since the initial emergence of the MFR, the formation of a linear bright current sheet between the flare loop and the filament, shown in Fig. \ref{fig:theSun}(a), was clearly observed by the EUV Imaging Spectrometer (EIS)/Hinode \citep{warren2018spectroscopic}, the Atmospheric Imaging Assembly \citep[AIA,][]{lemen2011}/Solar Dynamics Observatory (SDO) \citep{li2018spectroscopic} and also by the Solar Ultraviolet Imager (SUVI) on the GOES-16 spacecraft \citep{seaton2018observations}. 
The high-resolution imaging and spectroscopic observations show that the current sheet had a very high temperature ($>$10 MK) and very large nonthermal velocities ($>$150 km/s)\replaced{ and}{. It also} exhibited turbulent features associated with cascading magnetic reconnection process \citep{li2018spectroscopic} which were likely responsible for the initial stage of the acceleration of particles.  
Highly energetic particles have also caused hard X-ray emissions in the flare (via bremsstrahlung radiation) observed up to at least 300 keV by the Reuven Ramaty High Energy Solar Spectroscopic Imager \citep[RHESSI,][]{lin2002reuven} and the Fermi Gamma-ray space telescope, with two broad X-ray bursts centered at 15:57 and 16:10 UT on 2017-09-10, respectively.

The launch of the extremely fast erupting MFR and CME likely drove a global shock ahead of it indicated by the extreme-ultraviolet (EUV) waves \citep[{or EIT waves,}][]{dere1997} {in SUVI's 195 \r{A} passband (Fig.\ref{fig:theSun}(b) and better shown in a movie in the online version of \citet{seaton2018observations}). This strong EUV wave} had a speed of at least 1000 km/s, which places it amongst the fastest EUV waves observed \citep{long2018plasma}\replaced{, and}{. It} propagated across the entire solar disk within half an hour starting from the eruption at 15:53 \citep{seaton2018observations}.
This could indicate a global shock propagation within this short time period \citep{long2017understanding}. 
In fact, a signature of shock-related type II like radio emission (detected by the Greenland radio monitor) started at around 15:53\replaced{ and almost}{. Almost} simultaneously, type III radio emission (related to keV nonthermal electrons propagating outwards along magnetic field lines) was detected by the STA WAVES instrument (the WIND spacecraft at Earth did not have observations at this time period) suggesting the initial release of accelerated particles. 

Starting from 16:15 UT on {2017-09-10}, solar energetic particles (SEPs) arriving at Earth were registered as a ground level enhancement (GLE) seen by multiple neutron monitors with cutoff rigidities up to about 3 GV (corresponding to 2 GeV protons) as shown in Fig. \ref{fig:insitu}(b).
Different energy channels in GOES (panel (a)) clearly show an intense, sudden and long-lasting enhancement of the accelerated protons with energies larger than hundreds of MeV. %Considering 200 MeV protons streaming along the nominal interplanetary magnetic field (IMF) or Parker spiral (the magnetic connection from the onset to Earth is very good as shown in Fig. \ref{fig:dbm}(a)), it would take between 16-18 minutes for protons to propagate along the IMF of 1.1-1.2 AU (depending on the solar wind speed which was around 400-500 km/s). This indicates that such particles were likely released between 15:57 and 15:59 UT. 
From the clear onset time of relativistic particles, a release time around {16:00 UT} can be inferred {for 1 GeV protons}. 
This timing matches reasonably well with the final eruption of the flux rope and the X-ray bursts. 
However the timing of the initial signature of the shock and the reconnection process was very close (both starting around 15:53) and it is difficult to tell whether the shock or magnetic reconnection (flare) {contributed more to} the initial acceleration of particles. It is likely due to the combination of both as often observed in such eruptive and complex events \citep[e.g.,][]{aschwanden2002particle}.

\subsection{The early kinematics of three CMEs launched from September 9th to 10th}\label{sec:GCS}
To further track the erupted MFR and CME propagation into the interplanetary space, it is important to understand the contextual solar and heliospheric conditions prior to this event. Starting {from 2017-09-09}, two CMEs were seen in {the STA and the SOHO coronagraph} images {as shown in Fig. \ref{fig:theSun}(c)-(d)}. The two CMEs (named CME1 and CME2 in the order of their launch sequence) were launched before the aforementioned CME on {2017-09-10} (named CME3) {from} the same AR with similar directions. 

We utilized the graduated cylindrical shell (GCS) model \citep{Thernisien2009, thernisien2011} based on stereoscopic coronagraph observations of STA and SOHO to reconstruct the initial 3D geometry and kinematics of the CMEs. 
The GCS fits of CME1, CME2 and CME3 are shown in Fig. \ref{fig:theSun} (c), (d) and (e) respectively. % as green/yellow dashed contours over-plotted on the WL images.
{The northern and ecliptic components of CME2 have been reconstructed separately as the idealized single GCS reconstruction is not sufficient to describe the asymmetrical structure of CME2. However, only the component in the ecliptic plane was used as input for the later kinematics and propagation of CME2 in the interplanetary space (Table 1) as modeled by the 2D drag-based model (Section \ref{sec:DBM-cme}). In the coronagraph images of CME3, the flux rope is distinguished as a bright and structured component while the associated shock front is identified as a fainter quasi-spherical feature ahead of it. }
 
Multiple GCS fits over different time steps {were} used to derive the CME kinematic evolution. Height-time and velocity-time profiles of the CME apex are given in Fig. \ref{fig:theSun}(f) which show that {CME1 and 2} had moderate {and roughly constant} launch speeds while {CME3 erupted extremely rapidly} ($>$ 2600 km/s at the apex) {and was subsequently decelerated}. 
This is consistent with the trajectory and velocity of the flux rope of CME3 below 2 solar radii ($R_s$) derived by \citet{seaton2018observations} where its velocity was approximately 2000 km/s at 1.5 $R_s$, which suggests that the CME continued to accelerate up to a few $R_s$ (our GCS fitting started from $\approx$ 3 $R_s$).%This speed is smaller than our values mainly because the acceleration of the CME likely continued during its eruption upto a few $R_s$ (our GCS fitting started from $\approx$ 3 $R_s$) suggested by statistical studies of the acceleration process of CMEs \citep{Bein2011}. %and maybe also because their analysis neglected the projection effect of the event on the limb.

CME3 {drove} a globally propagating shock wave, {observed in the low corona} as {an} EUV wave {as discussed in Section \ref{sec:flare}}. {We reconstructed the CME3 shock kinematics focusing on its initial velocity using the GCS model }. The shock was modeled as a sphere-like structure with one pole attached to the solar surface (Fig. \ref{fig:theSun} (e)). {Although this assumption {does} not match well the observations which quickly extended into a global structure, the front part of the shock in the ecliptic plane} can be fitted well with GCS, from which we derived the initial shock speed {as also shown in Fig. \ref{fig:theSun} (e) together with the velocities of three CMEs}.
We assume the direction of the shock to be the same as {that of} CME3 and, as will be shown later, we fine-tune its geometric extent based on {available \textit{in-situ} plasma observations.} % {the IMF connection with Mars} which triggered the GLE at Mars.
{The longitudinal direction of the central portion of each CME/shock in the Heliocentric Earth Ecliptic (HEE) coordinate, its longitudinal half-width and launch speed} derived using GCS reconstruction are listed in Table \ref{tab:CME_kinematics} and also illustrated in Fig. \ref{fig:dbm}(a)-(b). 
{The plane-of-sky mass} of each CME was estimated based on SOHO/LASCO C2 images \citep{colaninno2009first} and their approximate values are also listed in the table. %and may have been slightly underestimated since the line of sight effect was neglected. CME1: 3.4(09/09 22:06) CME2: 3.5 (10/09 00:42) CME3 9.1 (10/09 16:30)

\begin{table}[ht!]
	\caption{The mass and GCS reconstructed initial kinematics of three CMEs and CME3-driven shock as well as the launch information (direction, location (distance from the Sun), time, and speed) and drag-parameter $\gamma$ for DBM.}\label{tab:CME_kinematics}
	\centering
	\begin{tabular}{ l|c|c|c||c|c|c|c } %{l|cccccc}
		\hline
		 & long. & 1/2 width & mass & DBM launch & time & speed & $\gamma$ [10$^{-7}$ \\
		 & HEE & [degree] & [$10^{15}$ g] & direction/location & dd/mm hh:min & [km/s] & km$^{-1}$]\\
		\hline
		CME1 & 119 & 35 & 3.4 & apex/20 $R_s$ & 09/09 23:46 & 500 & 0.1 \\
		CME2 & 116 & 19 & 3.5 & apex/20 $R_s$ & 10/09 02:16 & 1000 & 0.05 \\
		CME3 & 110 & 67 & 9.1 & apex/17.6 $R_s$ & 10/09 16:54 & 2600 & 0.01 \\
		\hline
		CME1+2 & 119 & 35 & $2M_0$$^{a}$ & to Mars/24 $R_s$ & 10/09 04:50 & 750 & 0.05 \\
		CME1+2+3 & 110 & 67 & $5M_0$$^{a}$& to Mars/68 $R_s$ & 10/09 21:00 & 1800 & 0.052\\
		\hhline{=|=|=|=|=|=|=|=}
		\multirow{2}{*}{shock} & \multirow{2}{*}{110} & \multirow{2}{*}{110-122$^{b}$} & \multirow{2}{*}{NA}& to Mars/18.1 $R_s$ & 10/09 16:54 & 2500 & 0.15\\
    & & & & to Earth/11.6 $R_s$ & 10/09 16:54 & 1600 & 0.4\\
		\hline	
		\multicolumn{8}{l}{$^{a}$ The mass of CME1, CME2 and CME3 was approximated as $M_0$, $M_0$ and $3M_0$ where $M_0 \approx 3 \times 10^{15}$g} \\
		\multicolumn{8}{l}{as only their mass ratio matters for the interaction kinematics treated in DBM \citep{temmer2012characteristics}.} \\
		\multicolumn{8}{l}{$^{b}$ The half-width of the shock in the interplanetary space is given in a range constrained by \textit{in-situ}}\\
		\multicolumn{8}{l}{observations (see Section \ref{sec:DBM-shock}).}\\
		\end{tabular}
\end{table}

\section{The interplanetary trajectory and interaction of 3 CMEs modeled by the drag-based model}\label{sec:DBM-cme}

Assuming that the main force that governs the propagation behavior of a CME in interplanetary space is the magnetohydrodynamical drag force, we simulated the kinematic profile of the CMEs via the drag-based model \citep[DBM, ][]{vrsnak2007transit, vrvsnak2013propagation}. 
We used the 2D DBM with the leading edge of the CME considered to be a semi-circle (diameter is the CME full angular width) such that although the apex initially propagates faster compared to the flanks, the variation of speed along the CME front decreases in time and the front gradually flattens during the propagation \citep{zic2015heliospheric, dumbovic2018drag}.
%Initial extents of 3 CMEs derived from GCS reconstructions are shown in Fig. \ref{fig:dbm}(a).

Within one day the three CMEs erupted in a similar direction, {each with a higher speed than the preceeding one,} hence, we expect that CME2 catches up and interacts with CME1 and later on CME3 catches up and interacts with the previous 2 CMEs. 
%In-situ observation (more details later) of the merged interplanetary CME (ICME) was not good enough to resolve its internal structure and it is impossible to tell if the CME MFRs fully merged after the interactions. 
%However, we could 
We assume that their mass merged as an entity and the two colliding bodies continued their propagation further with the momentum conserved before and after the interaction \citep{temmer2012characteristics}. This assumption is supported by \citet{shen2012super} who also suggested that the influence of CME kinematics by solar wind is much smaller compared to that due to collisions. After each CME-CME interaction in DBM we re-launched the merged CME from the merging location with re-evaluated drag parameters $\gamma$ until the next interaction. 

We used the solar wind speed of 500 km/s which was the average \textit{in-situ} measurement before the ICME/shock arrival at Mars/Earth (shown in Fig. \ref{fig:insitu}) and was also constrained by the shock propagation discussed later in Section \ref{sec:DBM-shock}. The drag parameter $\gamma$ used in the DBM for each CME before and after the interactions is shown in Table \ref{tab:CME_kinematics}. Since $\gamma$ depends on the CME mass and cross-sectional area, it was re-calculated after each interaction \citep{temmer2012characteristics}.
The initial $\gamma$ of {the} three CMEs were empirically set to decrease after one another since earlier CMEs have been observed to be able to efficiently 'sweep the way' and decrease the drag force for successive CMEs \citep{temmer2015interplanetary}. 
The choice of $\gamma$ and solar wind speed has been fine-tuned, through a forward modeling process using different input parameters, to best match the \textit{in-situ} arrival of the merged ICME at Mars %, which is characterized by the large Forbush decrease starting from 02:50 UT on 13/09 
{marked by the magenta bar in Fig. \ref{fig:insitu}(f)-(h)} (the ICME ejecta did not reach operational spacecraft at Earth and other locations).
 
As shown in Fig. \ref{fig:dbm}(g) and Table \ref{tab:CME_kinematics} of the results from DBM, \deleted{given the differences of their launch speeds,} CME2 caught up with CME1 at about 24 $R_s$ at 04:50 UT on {2017-09-10}\replaced{and the}{. The} merged CME (named CME1+2) had a cross-section combining the 2 CMEs which is equal to CME1 as it is wider than CME2 on both edges\replaced{ and}{. CME1+2 had} a speed of 750 km/s based on momentum conservation \added{before and after the collision}. 
The entity was later caught up by CME3 at about 68 $R_s$ at 21:00 UT on {2017-09-10} and the merged CME1+2+3 had a width of CME3 and a speed of $\approx$ 1800 km/s. It arrived at Mars at about 08:20 UT on {2017-09-13 (Fig. \ref{fig:dbm}(f) and (g))} with an arrival speed of about 748 km/s\replaced{which}{. This} is comparable to the Mars-EXpress (MEX) measurement {by the ASPERA-3 instrument \citep{barabash2006analyzer}} in the solar wind, which is however very scarce (black squares in Fig. \ref{fig:insitu}(h)). We will discuss about the modeled ICME and its shock arrival in comparison with \textit{in-situ} observations in Section \ref{sec:mars-model-obs}.

\section{Shock kinematics and propagation towards Earth and Mars: data constrained drag-based model}\label{sec:DBM-shock}

The fast and global propagation of the EUV wave discussed in Section \ref{sec:Sun} indicates a wide extent of the shock {reaching the direction of Earth}. \textit{In-situ} measurements at Earth clearly reveal the shock arrival as shown in Fig.\ref{fig:insitu}(c)-(e) with the 5-min resolution OMNI data. % which combines solar wind measurement from different spacecraft (as extracted from NASA/GSFC's OMNI data set through OMNIWeb).
The magenta bars in (a)-(e) mark the shock arrival at Earth characterized by enhancements of the magnetic fields, solar wind velocity, density, temperature and plasma flow pressure {as well as the Forbush decrease (FD) in various neutron monitors and GOES high-energy particle fluxes. FDs are identified as temporary and rapid depressions in the GCR intensity caused by interplanetary disturbances and magnetic shielding against charged particles during the passage of shocks and/or magnetic clouds \citep[e.g.][]{cane2000}.}

Towards Mars, the left flank of the ICME shock and ejecta are expected to hit the planet. 
{Unfortunately, \textit{in-situ} solar wind and magnetic field observations at Mars upon the ICME shock arrival are very limited as shown in Fig. \ref{fig:insitu}(h). A clear signature indicating the shock arrival is the FD at $\sim$ 02:50 UT on 2017-09-13 detected by the Radiation Assessment Detector \citep[RAD,][]{hassler2012} onboard the Mars Science Laboratory (MSL) rover on the surface of Mars. Compared to previous FD observations at Mars \citep{guo2018measurements}, this event has a magnitude \replaced{larger than 20\%}{up to $\sim$ 23\%} in the RAD dose rate which makes it the largest FD observed by RAD since the landing of MSL in August 2012}.
\added{\citet{witasse2017interplanetary} have studied one of the largest FD event seen by RAD with a $\sim$ 19\% magnitude of decrease observed on 17 October 2014. This event is similar to the 2017 September event studied here that Mars were located at the east flank of both CMEs. 
But the launch speed of the 17 October 2014 CME was much smaller $\sim$ 1015 km/s while the September 2017 event had a launch speed of about $2600$ km/s (Table \ref{tab:CME_kinematics}). 
Consequently, the transit time of the September CME from the Sun to Mars is only about 57.5 hours which is $\sim$ 11 hours shorter than the 17 October 2014 event.}
%\citet{witasse2017interplanetary, guo2018measurements, von2018using} describe the first and recent observations of FDs at Mars.

%\replaced{The longitudinal extent of the shock has been constrained with the following considerations. Assuming the shock extent is symmetric and its central longitude having the same direction as CME3 (110$^\circ$ in HEE coordinate) and given the right edge of the shock passing Earth as observed \textit{in-situ}, the left-edge of the shock should reach longitudes $\ge$ 220$^\circ$ {(Fig. \ref{fig:dbm}(b))}. However, the shock should not be much wider, as otherwise it would reach STA (232$^\circ$) which is not supported by \textit{in-situ} observations shown in Fig. \ref{fig:insitu}(i)-(k).}
{We constrained the longitudinal extent of the shock based on 1) the assumption that the shock is symmetric around the direction of CME3 (110$^\circ$ in HEE coordinate), 2) the \textit{in-situ} OMNI data showing that the right edge of the shock passed Earth and 3) STA plasma and magnetic field measurements %\citep{acuna2008stereo, galvin2008PLASTIC} 
suggesting that the left-edge of the shock should not reach STA at 232$^\circ$ (Fig. \ref{fig:insitu}(i)-(k)).} 
This constrained half-width of the shock is {between 110 and 122 degrees as} given in Table \ref{tab:CME_kinematics} and Fig. \ref{fig:dbm}(b).

%\replaced{To better derive the propagation kinematics of the shock, we used 1D DBM (without any shock geometry) to track the shock front in Earth and Mars directions separately. The parameters are also allowed to differ in Earth and Mars directions which are $\sim$ 160$^\circ$ apart in longitude.}
{Because Earth and Mars were $\sim$ 160$^\circ$ apart, different interplanetary conditions should be considered for the shock propagation in each direction.} {Various} solar wind speeds and drag parameters $\gamma$ were tested for multiple runs and for each run, we compared the modeled results with \textit{in-situ} observational constraints including: 
\begin{itemize}
\item Shock arrival time at Mars should be around 02:50 UT on 2017-09-13 which corresponds to the onset of the FD seen by MSL/RAD on the surface of Mars as shown in Fig. \ref{fig:insitu}(f)-(g). 
\item The shock arrival speed at Mars should match the \textit{in-situ} solar wind speed which is about 650-800 km/s indicated by the scarce but precious solar wind data measured by MEX {(Fig. \ref{fig:insitu}(h))}. 
\item Shock arrival time at Earth should be around 18:30 UT on 2017-09-12 which is suggested by magnetic field and solar wind measurement at Earth (Fig.\ref{fig:insitu}(a)-(e), {magenta bars}).
\item The shock arrival speed at Earth should match the \textit{in-situ} solar wind speed which is about 600 km/s shown in Fig.\ref{fig:insitu}(c). 
\item The solar wind speed prior to the shock arrival at Earth/Mars {varied between 400-600 km/s} as shown in Fig. \ref{fig:insitu}(c)/(h). Different DBM runs {with} 400, 500 and 600 km/s were performed and compared.
\end{itemize}
The optimized fitting result {from these} DBM runs is shown in Fig. \ref{fig:dbm}(f) where the shock was launched with different speeds towards Earth and Mars as derived from GCS fits (Table \ref{tab:CME_kinematics}). The best-derived $\gamma$ values are 0.15 and 0.4$\times10^{-7}$ km$^{-1}$ in the direction of Mars and Earth respectively while the best-matching solar wind speed is 500 km/s in both directions. 
We note, in order to match the shock arrival time at Earth and Mars we had to use rather large $\gamma$ values compared to those for the associated ICME. 
We justify the increased drag by the assumption that the shock caused by CME3 is only weakly driven over certain distance ranges. In the Mars direction, this happens beyond the distance of 68 $R_s$ due the sudden deceleration of CME3 as it interacts with the previous CMEs. Towards Earth, the shock is even less strongly driven as the main propagation direction of the magnetic structure is directed towards Mars, and, hence, experiences a larger drag.
%We note that in oder to match the arrival time at Earth the derived $\gamma$ is 0.4$\times 10^{-7}$ km$^{-1}$ is unusually high compared to empirical values \citep{vrvsnak2014heliospheric}. 
%We note that in order to match the arrival time at Earth and Mars the derived values of $\gamma$ are unusually large compared to the values of $\gamma$ for its {associated} ICME. This is because the shock was driven and relied on CME3 at first and CME 1+2+3 after the interaction. 
%In our modeling of the shock propagation, the value of $\gamma$ must be large to compensate for the sudden deceleration of CME3 during its interaction with the previous CMEs. 
%Besides, $\gamma$ for the shock toward Earth is even much higher than that for the shock toward Mars. This probably reflects that the shock in Earth direction was not strongly driven by the expanding magnetic structure, hence, experiences a larger drag.
%However, DBM is not intended for modeling shock kinematics \citep{vrvsnak2013propagation, liu13}, and the high $\gamma$ value may reflect that the shock in Earth direction was not strongly driven by the expanding magnetic structure, hence, experiences a larger drag. 

\section{The shock and ICME arrival at Mars and Earth: modeled results and \textit{in-situ} observations}\label{sec:mars-model-obs}
%\textit{In-situ} observations at Mars of the ICME structure are very limited and we refer the readers to Lee et al (this issue) for more discussions of the ICME arrival identified in the measurements of the Mars Atmosphere and Volatile Evolution Mission (MAVEN). Here
\textit{In-situ} observations at Mars of the ICME structure are very limited (Fig. \ref{fig:insitu}(h)) and the magenta highlighted bars in (f)-(h) mark the possible passage of the shock and ICME at Mars indicated by MEX measurement in the solar wind {overplotted with the WSA-ENLIL \citep[][and references therein]{Mays2015} modeling results. The current run of the WSA-ENLIL plus cone model (run ID 'Leila\_Mays\_120817\_SH\_9' on the CCMC server) has included the launch and propagation of the aforementioned three CMEs and is explained in better details in \citet{luhmann2018shock}}. 
Note that similar to DBM, input parameters for the ENLIL modeling were tweaked in order to best match the observations. Unlike the decoupled structures in DBM, CMEs in ENLIL could drive the shock front in a more physical manner. {The ENLIL modeled ICME shock arrived at Mars at about 04:00 UT on 2017-09-13 which is very close to the onset time of the FD at Mars (Fig. \ref{fig:insitu}(g)) and the solar wind speed and density peaked at around 820 km/s and 3.2 protons/cm$^3$ which are consistent with the MEX measurement during the ICME passage (Fig. \ref{fig:insitu}(h))}.
 
{Given the direction and the longitudinal extent of the three CMEs derived from the GCS model (Table \ref{tab:CME_kinematics} and Fig. \ref{fig:dbm} (a)-(b)) and the \textit{in-situ} observation at Earth (Fig. \ref{fig:insitu}(c)-(e)), no ICME ejecta but only the right flank of the shock arrived at Earth. 
The ENLIL modeled shock arrived at Earth at $\sim$ 00:00 UT on 2017-09-13 which is about 5.5 hours later than the \textit{in-situ} detection of the shock arrival \added{(Fig. \ref{fig:insitu}(c)-(e))}. The modeled peak magnetic field strength and solar wind speed are also slightly smaller than the measured values. 
Considering the complexity of the events, the shock/ICME arrival at both Earth and Mars modeled by ENLIL matches reasonably well with observations within the limit of statistical uncertainties.  
\added{The mean absolute arrival-time prediction error was about 12 hours as studied by \citet{Mays2015} of 17 CMEs which were predicted to arrive at Earth.} %{Taktakishvili2010}. 

{The DBM modeled results of the shock arrivals at Earth and Mars are illustrated in Fig. \ref{fig:dbm} (f). 
The launch speed of the shock in the direction of Earth was slightly smaller than that towards Mars as derived from GCS fits (Table 1). The drag parameter is also larger for the shock propagating towards Earth as it was not really driven by the ICME magnetic structure in this direction. 
The DBM predicted shock arrival at Earth is at about 18:15 UT on 2017-09-12 with an arrival speed of $\sim$ 625 km/s which are very close to the observational arrival time of 18:30 and speed of $\sim$ 630 km/s (which is expected as DBM is tuned to match the observation). 
The modeled shock arrived at Mars with a speed of 775 km/s at around 02:47 UT on 2017-09-13 which is perfectly matching the onset time of the RAD seen FD at 02:50 as shown in Fig. \ref{fig:dbm} (c) .}

{As modeled by the DBM and described in Section \ref{sec:DBM-cme}, three CMEs were launched in similar directions and interacted with one another as they propagated towards the direction of Mars. 
The merged entity arrived at Mars at about 08:20 UT on 2017-09-13 (Fig. \ref{fig:dbm}(g)) with an arrival speed of about 748 km/s which agrees with the MEX solar wind speed of 714 km/s measured hours later at 22:57 UT. 
Unfortunately, \deleted{no }\textit{in-situ} interplanetary magnetic field (IMF) or solar wind data are \replaced{available}{rather limited} for identifying the arrival and the structure of the magnetic ejecta {\citep{lee2018maven}}.}

Upon the ICME's arrival at Mars, the FD measured by RAD on the surface of Mars had a decrease in the high-energy count rate up to $\sim$ 23\% (Fig. \ref{fig:dbm}(c)) and is the biggest FD detected by RAD to date.
A classical picture of FDs \citep{cane2000} suggests that an ICME with a shock front passing by an observation point could result in a 2-step structure, i.e., with the 1st decrease corresponding to the shock arrival and the turbulent sheath region and the 2nd step indicating {passage} of the magnetic ejecta. 
However, recent studies suggest that the ejecta may not always be associated with a decrease, especially at \replaced{larger distances to}{distances further away from} the Sun \citep{winslow2018opening}.
As shown in Fig. \ref{fig:dbm} (c) and (f), the modeled shock arrival time at Mars agrees nicely with the initial FD onset while the ICME (merged ejecta) arrival might {correspond} to a rather weak second decrease. 
However, due to the scarce \textit{in-situ} magnetic and plasma measurement in the solar wind, we can not confirm the 2-step FD profile and its corresponding ICME structure. 

\section{SEPs arriving at Earth, Mars and STA and the indication of the shock and SIR propagation}\label{sec:SEPs}
As discussed in Section \ref{sec:flare}, the onset of relativistic particles at Earth was about 10-15 minutes after the flare onset indicating a good magnetic connection between the particle injection site and Earth. 
As shown in Fig. \ref{fig:insitu}(j), high energetic protons \added{started }slowly arriving at STA %\citep{von2008high}
\added{at around $\sim$ 08:00UT on 2017-09-11 which was} \replaced{had $\sim$ 1 day of delay}{$\sim$ 16 hours later than the flare onset.} \replaced{  which}{The arrival of these SEPs} is probably attributed to cross-field diffusion in the solar wind \citep[e.g.,][]{Droge2010}. 
{This is supported by Fig. \ref{fig:dbm} (a) which shows that STA was magnetically connected\added{, under various different solar wind speeds,} to the back side of the Sun where the flare erupted.}
At Mars, the earliest possible onset of $>100$ MeV protons is at 19:50 UT, $\sim$ 215 minutes later than that at Earth. \added{Considering the Mars IMF footpoint separation from the fare longitude is about 135 degrees, this onset delay is within the statistical uncertainties of high energy proton onset delay studied by \citet{richardson2014} (Fig. 16).} However, it is unclear whether this delay is due to cross-field diffusion or a later magnetic connection to the acceleration/injection site or both. 

First we consider the model with continuous particle injection at the shock as it propagates outwards {(due to re-acceleration of particles by the interplanetary shock)} and establishes magnetic connection to {the observer} \citep[e.g.,][]{lario2013longitudinal, lario2017solar}. 
Given the proton onset at {19:50 seen by the highest energy channel of MSL/RAD} %and electron onset at 19:20 at Mars (private communication with the MAVEN/SEP team), 
this model requires {that} the shock started connecting to the Parker spiral towards Mars under a solar wind speed of 500 km/s at $\sim$ {19:30} UT to allow for parallel and efficient particle transport to Mars. 
As modeled by DBM (Fig.\ref{fig:dbm}(f)), at $\sim$ {19:30}, the shock front in the direction of Mars has a propagation distance of {47} $R_s$ %under a 500 km/s solar wind speed 
and {the magnetic establishment} would require the shock to have a half-width of about {114}$^{\circ}$ (cone boundaries shown as yellow lines in Fig.\ref{fig:dbm}(a)-(b)) which is in-between the constrained range (Section \ref{sec:DBM-shock} and Table \ref{tab:CME_kinematics}). {The path of particles along the 500 km/s Parker spiral from the shock front to Mars is highlighted in red.} 
{We note that the solar wind speed of 500 km/s is an approximation of the observation and a fitted parameter from DBM. With a slightly faster solar wind speed (e.g., the 600 km/s IMF plotted as dotted lines in Fig. \ref{fig:dbm}(a)/(b)), the magnetic establishment at $\sim$ {19:30} requires a smaller shock width. 
Alternatively, if the solar wind speed is about 400 km/s (solid curves in the plots), it has to be considerably wider to establish the magnetic connection upon the SEP onset under the condition of \deleted{the }undisturbed IMF. However, this wider shock, while propagating radially outwards, should also reach STA which is however not supported by the STA \textit{in-situ} observations (Fig. \ref{fig:insitu}(i)-(k)).}

In the scenario of continuous {particle injection at the shock front}, the preceding 2 CMEs may provide a "seed" population for the catching-up shock \citep{Gopalswamy2002, Gopalswamy2004}. 
In fact, a small jump in the GOES data at around 21:20 UT as indicated by the vertical red line in Fig.\ref{fig:dbm}(d) may indicate the injection of more particles at the shock through merging of CME3 with CME1+2 at around 21:00 UT {predicted by DBM} (Table \ref{tab:CME_kinematics} and Fig. \ref{fig:dbm}(g)). %Particle measurement at Mars is however not good enough for such an identification. 
Alternatively, as the Earth connection point along the shock front changes, the discontinuity or evolution of the shock parameters may also contribute to the second peak as observed in-situ at around 21:20.

%However, the above model assumes the shock to be the main contributor of SEPs and excludes any perpendicular diffusion process that may have started near the solar surface. 
Nevertheless, particle scattering and transport across the IMF could have also played a role during the event. As STA was not magnetically connected to the flare or the shock (Fig. \ref{fig:dbm}(a)) {from the beginning}, early SEPs detected at STA {(gradual time profile of the flux)} should have been transported there across IMF lines.
Towards the direction of Mars, with a smaller width of 110$^\circ$ ({left edge of the shock is marked in cyan in} Fig.\ref{fig:insitu}(b)) which is the lower limit of the constrained shock width, the DBM modeled shock could not be connected to the 500 km/s Parker spiral towards Mars upon the SEP onset. 
In this case, SEPs first arriving at Mars are likely due to particle transport across IMF {from the injection site which could be the shock front and/or the flare reconnection region}. 

%{At about 09:10 UT on 2017-09-12, STA shows an enhanced bump in the flux of SEPs $\le$ 40 MeV (Fig.\ref{fig:insitu}(i)) and at this hour the shock propagation in the direction towards Mars has a distance of about 1.2 AU as modeled by DBM. 
%With a shock width of $\sim$ 110$^\circ$, this would trigger a direct magnetic connection of the left edge of the shock back to STA which could possibly explain the the enhanced SEP flux at this time. } 

We have compared the \textit{in-situ} observations with the results from the WSA-ENLIL predictions at Mars, Earth and STA in Fig. \ref{fig:insitu} (ENLIL results are plotted as dashed lines) {as discussed in Section \ref{sec:mars-model-obs}}. 
{From ENLIL simulations,} the shock information and propagation along the IMF passing certain observers could be extracted for each CME \citep{bain2016shock, luhmann2017modeling}. 
Extracted shock information in the current run indicates that the shock started connecting to the IMF towards Mars at about 06:00 UT on 11/09, {$\sim$ 10} hours later than the SEP onset at Mars. 
In such a case, cross-field transport, presumably close to the Sun, must have dominated over establishment of a direct magnetic connection to the shock. However, a detailed investigation would require careful studies of the particle transport modeling, taking into account effects such as adiabatic cooling, focusing, turbulent scattering, pitch angle scattering, and cross-field diffusion \citep[e.g.,][]{zhang2009propagation, hu2017}. 

{At a later phase of the SEP event, particles were widely distributed across heliospheric longitudes $\geq$ 232$^\circ$ (or even wider as this is constrained by three observational locations), the interaction of SEPs with large-scale heliospheric structures is particularly interesting. 
Stream Interaction Regions \citep[SIR,][]{Burlaga1974} are interaction regions between the fast and slow solar wind characterized by sudden changes in the flow density, temperature and significant increase of solar wind as well as compressed magnetic field. Stable SIR structures may co-rotate with the Sun which has a rotation period of $\sim$ 27 days in the solar equatorial plane. 
In Fig. \ref{fig:insitu}, we identified two SIRs passing Earth and STA respectively (named SIR1 for the one passing Earth and SIR2 for the one across STA) as highlighted in cyan areas.}

{As shown in Fig. \ref{fig:insitu}(i)-(k), SIR2 arrived at STA at about 22:48 UT on 2017-09-13 and the high energy proton flux (up to 100 MeV) has a small enhancement which suggests SEPs may have leaked into, get trapped and/or re-accelerated in the SIR structure. 
Considering the SIR rotates with the Sun, we time-shift the SIR2 structure observed at STA back to 2017-09-10 at {19:30} UT ($\sim$ shortly before the SEP event at Mars). As illustrated in Fig. \ref{fig:sir}(a), SIR2 arrived at Mars at almost this time. 
In fact, in-situ solar wind and magnetic field observations were available during this period and an SIR was identified to have impacted Mars at 23:30 UT {\citep{lee2018maven}} which perfectly agrees with the time-shifted SIR2 from STA to Mars. 
Fig. \ref{fig:insitu}(h) also shows the evolution of proton temperature, density and solar wind speed (from $\sim$ 300 to $\sim$ 500 km/s) at Mars during the SIR2 passage which is consistent with the solar wind changes when SIR2 passed STA.}

{Upon the SEP onset at Mars (Fig. \ref{fig:sir}(a)), SIR2 was connected even closer to the central part of the shock/flare than Mars. Therefore, SEPs were likely also injected into the SIR structure, preferentially along the IMF direction directly from the shock front as particles cannot easily penetrate through an SIR structure.  %or due to perpendicular diffusion across the IMF. 
%{\color{red}{[Can we argue then that there must be a direct magnetic connection of Mars to the shock front as the SIR is acting as a "barrier" for particles cross-field transport? Or, instead, SIR works as a tunnel for SEPs first arriving at Mars...?]}}
Such a scenario may have also contributed to the SEPs first arriving at Mars even if Mars were not directly connected to the injection site. 
These high-energy SEPs were trapped in (or perhaps even re-accelerated therein) and co-rotated with SIR2 and caused a remarkable enhancement of the SEP flux when SIR2 arrived at STA (Fig. \ref{fig:insitu}(i)) at 22:48 UT on 2017-09-13.
Since these SEPs were accelerated by the flare/shock closer to the Sun, they have a higher energy component (up to 100 MeV). 
%In comparison, at an earlier time of the STA particle flux, the enhanced bump at 09:10 UT on 2017-09-12 was only related with SEPs $\le$ 40 MeV as the shock was further away with less efficiency of particle acceleration.}

{On the other hand, SIR1 (cyan area in Fig. \ref{fig:insitu}(a)-(e)) had a more significant enhancement of the solar wind speed (Fig.\ref{fig:insitu}(c)) with a more compressed shock structure causing substantial FDs in the NM count rates. 
It passed Earth starting around 10:15 UT on 2017-09-14. 
Time-shift analysis shows that shortly after the flare and SEP onset, SIR1 was about 50$^\circ$ west of Earth and was barely magnetically connected to the right edge of the shock as shown in Fig. \ref{fig:sir}(a), thus making particle injection into SIR1 rather unlikely. 
It is evident in Fig. \ref{fig:insitu} (a) that between the shock structure (which passed Earth at 04:00 UT on 2019-09-13, magenta area) and SIR1 arrival at Earth, there is a plateau in the GOES high-energy flux which is different from the declining time profile before the shock arrived at Earth. 
This may be caused by energetic particles being trapped between the shock and SIR1, which act as two barriers for these SEPs, as illustrated in Fig. \ref{fig:sir}(b). %during this period 
{This is supported by \citet{strauss2016} who suggested that perpendicular diffusion could be strongly damped at magnetic discontinuities, which may be responsible for the large particle gradients associated with these structures such as an SIR.}
When SIR1 shock passed Earth, this reservoir of SEPs also passed Earth causing a sudden decrease of the GOES flux at energies below $\sim$ 80 MeV. This decrease is deferent from a normal FD in the GCR flux as seen by NMs on ground.}

\section{Summary and Conclusion}

We investigated and modeled the geometry, kinematics, propagation and interaction of three CMEs launched around 2017-09-10 from their solar origin to their arrivals at Mars and Earth.
The modeled results are constrained by and compared with in-situ measurements at Earth, Mars and STA. 
%The ICME arrival and its potential space weather impact at different heliospheric locations depend not only on the dynamic heliospheric conditions but also {on} the evolution of the ICME kinematics, especially during interactions of different CMEs.
Observation-based modeling of the ICME and the interplanetary shock reveals the complexity of the event and the advantage of more measurements for advancing space weather predictions.
The optimized modeling for the ICME arrival at both Earth and Mars suggests that in order to better predict the ICME arrival and its potential space weather impact at different heliospheric locations, it is important to consider 1) the evolution of the ICME kinematics, especially during interactions of different CMEs and 2) the dynamic heliospheric conditions at different locations in the heliosphere.

The SEP event associated with the flare and the eruption of the last CME has been detected, for the first time, at the surface of two planets, registered as GLE72 at Earth and the biggest GLE seen by MSL/RAD on Mars. 
{Relativistic particles first arriving at Earth and causing GLE72 were mainly accelerated by the flare and the initial shock.
The particle onset at Mars is $\sim$ 3.5 hours later than that at Earth and this was caused by either a later magnetic connection of Mars to the shock front which serves as an injection source for SEPs and/or cross-field diffusion of SEPs from the acceleration and injection site.
Particles started arriving at STA $\sim$ \replaced{1 day}{16 hours} later with a gradual rising profile indicating perpendicular diffusion across IMF was mostly responsible at this phase.  
%However an enhanced bump at around 09:10 UT on 2017-09-12 in SEP fluxes (energies $\le$ 40 MeV) may correspond to a direct connection to the shock which has a larger heliospheric distance than STA.
%Thus particles were transported back from the shock front to STA. Detailed analysis of the pitch angle information will be carried out in the future for better understanding the particle injection and transport towards STA. %{\color{red}{[We had a quick check of the anti-Sun and Sunward electrons which isn't too helpful...]} }
Numerical modeling of particle propagation in the heliosphere taking into account of the dynamic acceleration and injection process would also be helpful for understanding the interplanetary journey of these highly energetic particles arriving at three locations $>$ 230$^\circ$ longitudinally apart.}

{Two SIRs have been detected \textit{in-situ} at Earth and STA respectively. 
We shifted SIR2 (detected at STA) back in time and found that its arrival time at Mars is coincident with the SEP onset time at Mars and it had a magnetic connection even closer to the central part of the shock. 
This may have favored particles injected into the SIR which were later observed as an enhancement in the SEP flux when it passed STA. %Compared to the bump structure in the SEP flux, these are particles accelerated earlier closer to the Sun but detected later at STA. 
On the other hand, SIR1 arrived at Earth $\sim$ 1.5 days after the CME shock passed Earth and SEPs were trapped between these two structures causing a plateau profile in the GOES SEP flux. }

%
%Relativistic particles first arriving at Earth and causing GLE72 were mainly accelerated by the flare and the initial shock while later and continuous injection of SEPs at Earth, Mars and STA were more likely related to the widely-extending CME-driven interplanetary shock propagating outwards and reaching both planets as indicated by the time of shock connection with IMF towards Mars matching well the SEP onset time at Mars. 

\acknowledgments
We acknowledge use of NASA/SPDF OMNIWeb service and OMNI data, EU NMDB database (www.nmdb.eu), STEREO data (https://stereo-ssc.nascom.nasa.gov/) and NOAA GOES data (https://satdat.ngdc.noaa.gov/). ENLIL with Cone Model was developed by D. Odstrcil at George Mason University.
MSL RAD is supported by NASA (HEOMD) under JPL subcontract 1273039 to SWRI, and in Germany by DLR (under German Space Agency Grants 50QM0501, 50QM1201, and 50QM1701) to the Christian-Albrechts-University of Kiel.
RAD data are archived in the NASA planetary data systems' planetary plasma interactions node (http://ppi.pds.nasa.gov/).
The Swedish contribution to the ASPERA-3 experiment is supported by the Swedish National Space Board. ASPERA-3 data are public at the ESA Planetary Science Archive.
M.D. acknowledges funding from the EU H2020 MSCA grant agreement (745782). 
M.T. acknowledges the support by the FFG/ASAP Programme under grant number 859729 (SWAMI). 
Y.W. is supported by the grants from NSFC (41574165 and 41774178). 
A.M.V. acknowledges support from the Austrian Science Fund (FWF): P27292-N20.
J.G. thanks Christina Lee, Andreas Klassen, Fernando Carcaboso, Nina Dresing, Andreas Taut for helpful advices and discussions. 
J.G. and R.F.W.S. acknowledge discussions during various ISSI team meetings.

%\bibliography{msl_rad_guo}

\begin{figure}[ht!]
	\includegraphics[width=0.98\textwidth]{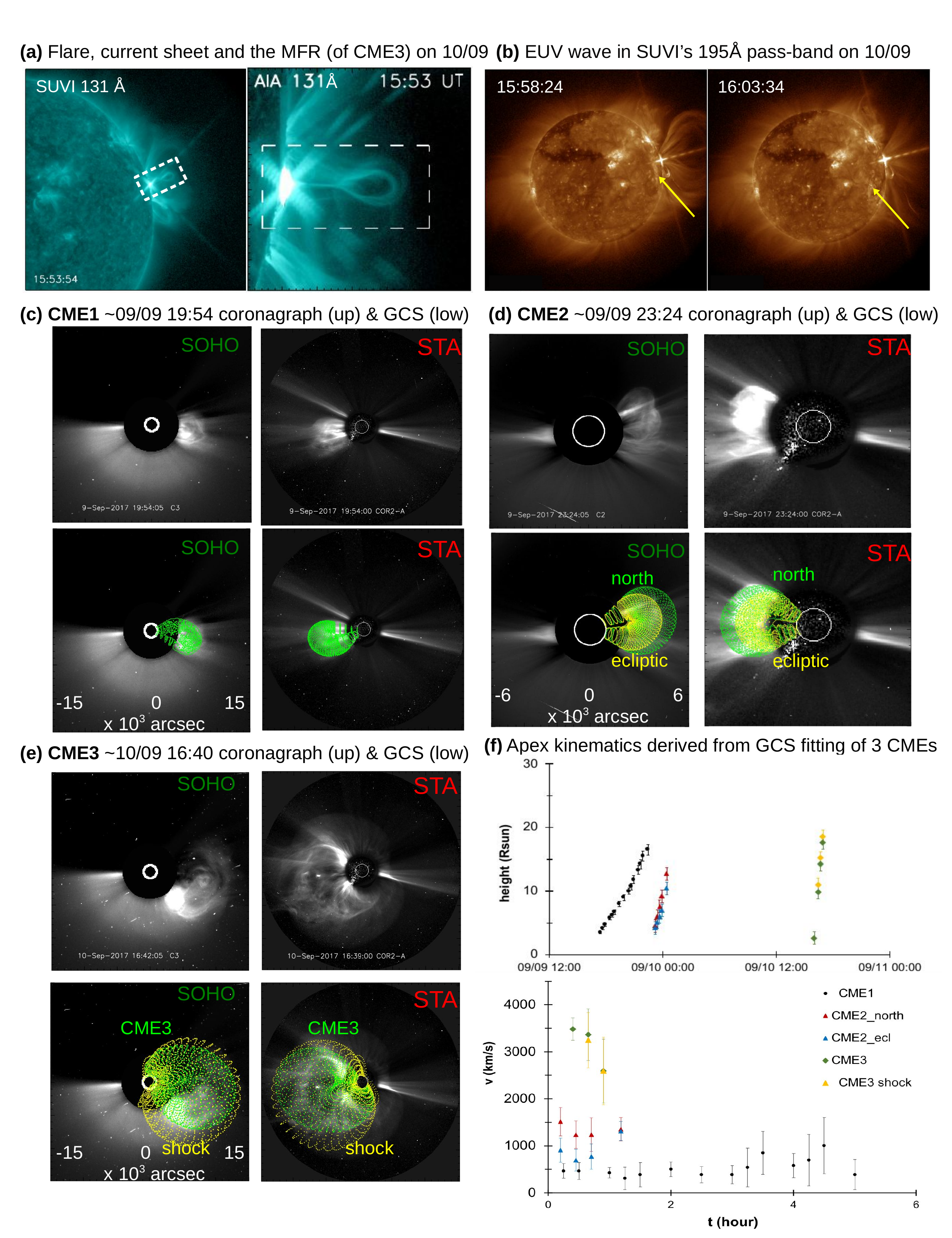}
	\caption{Remote sensing observations of the 2017-09-10 flare and three CMEs launched in the same AR from 2017-09-09 to 2017-09-10.
		(a) shows the GOES/SUVI 131 \r{A} (adapted from \citet{seaton2018observations}) and SDO/AIA 131 \r{A} (adapted from \citet{li2018spectroscopic}) observations of the flare and initial eruption of the MFR {(associated with CME3)} with the white dashed box marking the flare, current sheet and MFR.
		(b) displays the post-eruption phase of the MFR and fast propagation of the EUV wave away from the onset location with the yellow arrow pointing at the wave front. 
		(c)-(e) (top panels) show stereoscopic coronagraph WL images {(equal ranges in x and y axes)} of CME1, CME2 and CME3 at selected times. The bottom panels illustrate GCS reconstruction of the CME geometry. The northern and ecliptic components of CME2 have been fitted respectively and the CME and shock components of CME3 have also been modeled separately.
		(f) plots apex kinematics derived from GCS modeling of each CME evolution in time. The upper panel shows the height (in solar radii) versus time and the lower panel is the velocity-time plot where time is normalized to the the start of each CME. }
	\label{fig:theSun}
\end{figure}

\begin{sidewaysfigure}
	\centering
	% when using pdflatex, use pdf file:
	\includegraphics[trim=5 25 5 5,clip, width=0.95\textwidth]{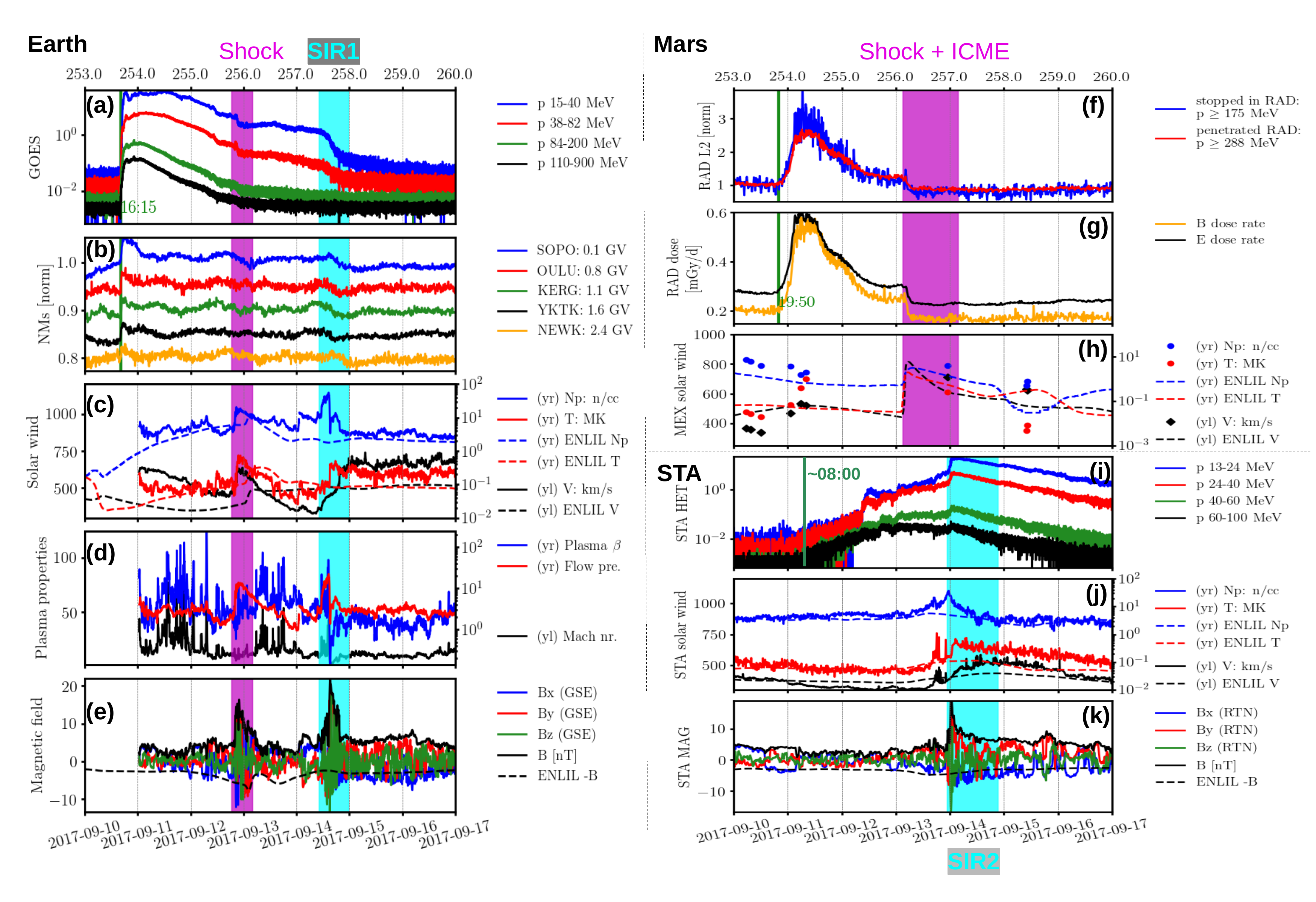}
	%
	% when using dvips, use .eps file:
	% \includegraphics[width=20pc]{figsamp.eps}
	%
	% If you don't specify the file extension, 
	% \includegraphics will insert the right file:
	%\centerline{\includegraphics[height=1.5in]{figsamp}}
	\caption{In-situ observations of the 2017-09-10 events at Earth (left panels), Mars (f-h) and STA (i-k). (a)/(i) shows the energy-dependent particle flux (counts/cm$^2$/sr/MeV/sec) measured by GOES and STA. (b) includes count rate of several ground-based neutron monitors with different cut-off rigidities (each neutron monitor data set is normalized to the average value of the selected time range and shifted apart in y-axis). (c) (h) and (j) present the solar wind speed (black, left y-axis (yl)), proton density (blue, right y-axis (yr)) and temperature (red, yr) for Earth (OMNI), Mars (MEX) and STA respectively. The ENLIL modeled results at three locations are also plotted as dashed lines. (d) plots the Alfv\'en Mach number (black, yl), plasma $\beta$ (blue, yr) and flow pressure (red, yr) estimated at Earth. (e)/(k) displays the vector magnetic fields at Earth/STA in Geocentric Solar Ecliptic (GSE) or spacecraft Radial-Tangential-Normal (RTN) coordinate and negative ENLIL modeled magnetic field strength (dashed line). (f) contains the normalized count rate for downward particles stopping in RAD and penetrating RAD with the former/later approximating protons with energies larger than 175/288 MeV arriving at Mars. (g) shows the dose rate recorded in RAD B (silicon) and E (plastic) detectors. {Magenta highlighted areas are the ICME and/or its associated shock passage at Earth and Mars.} Cyan highlighted areas are high-speed streams (of two different ones) passing Earth and STA during this period. Vertical solid lines in (a)-(b)/(f)-(g)/(i) indicate the particle onset time at Earth/Mars/STA.}
	\label{fig:insitu}
\end{sidewaysfigure}

\begin{figure}[ht!]
	\includegraphics[trim=5 4 0 8, clip, width=0.99\textwidth]{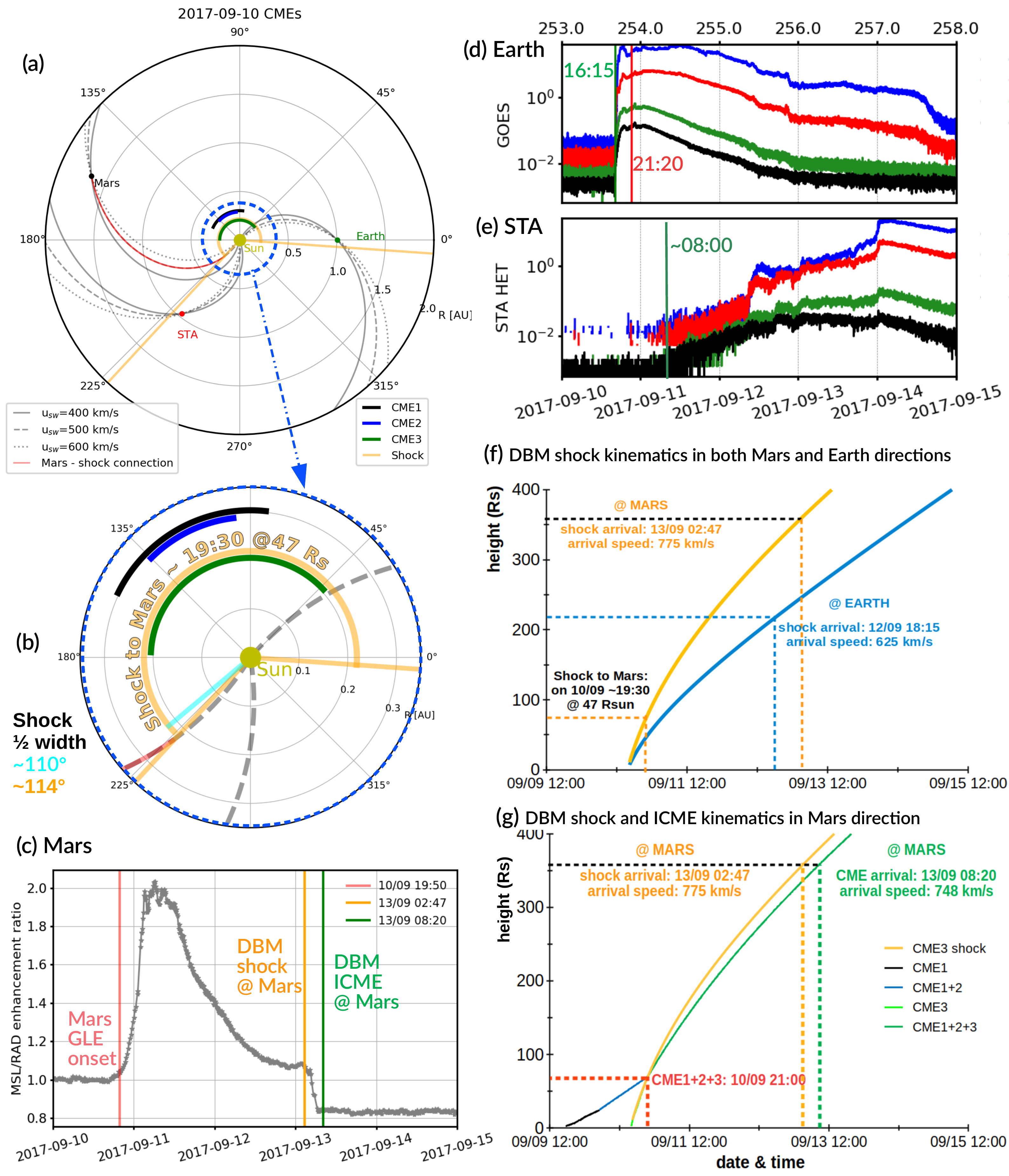}%trim=130 10 130 8, clip, 
	\caption{(a) shows heliospheric locations of the Earth, Mars and STA in HEE coordinates. Longitudinal extents of 3 CMES and the interplanetary shock driven by CME3 are approximated by circular contours (their geometries are not modeled as sun-centered circles). 
	Norminal IMFs passing three observers under different solar wind conditions are plotted. At $\sim$ {19:30} the shock (yellow countour) is at a distance of {47} $R_s$. 
	{Yellow cone boundaries show the extent of the shock with a half-width of {114}$^\circ$ which could connect to the 500km/s solar wind IMF towards Mars upon the SEP onset.}
	(b) is a zoom-in of (a) within 0.3 AU and also shows the left edge of the shock with a half-width of 110$^\circ$ not connecting to the 500km/s Parker spiral.
	(c) plots the {enhancement rate (to the background value) representing SEPs $\ge$ about 275 MeV reaching Mars. }%particles reaching RAD with energies $\ge$ 100 MeV. 
	(d) and (e) show the energy-dependent particle flux measured by GOES at Earth and STA (units and legends are the same as in Fig.\ref{fig:insitu}). 
	%The first SEP onset at Earth corresponds to the initial arrivals while the small bump may be related to CME3 merging with CME1+2. The SEPs arriving at STA are likely transported there across IMF lines as STA was not magnetically connected to the flare or the shock. 
	(f) and (g) are the best-fitting DBM results of the shock and ICME kinematics in Mars and Earth directions with more descriptions and discussions in Section \ref{sec:DBM-cme}-\ref{sec:mars-model-obs}. }
	\label{fig:dbm}
\end{figure}

\begin{figure}[ht!]
	\includegraphics[trim=40 80 40 60, clip, width=1\textwidth]{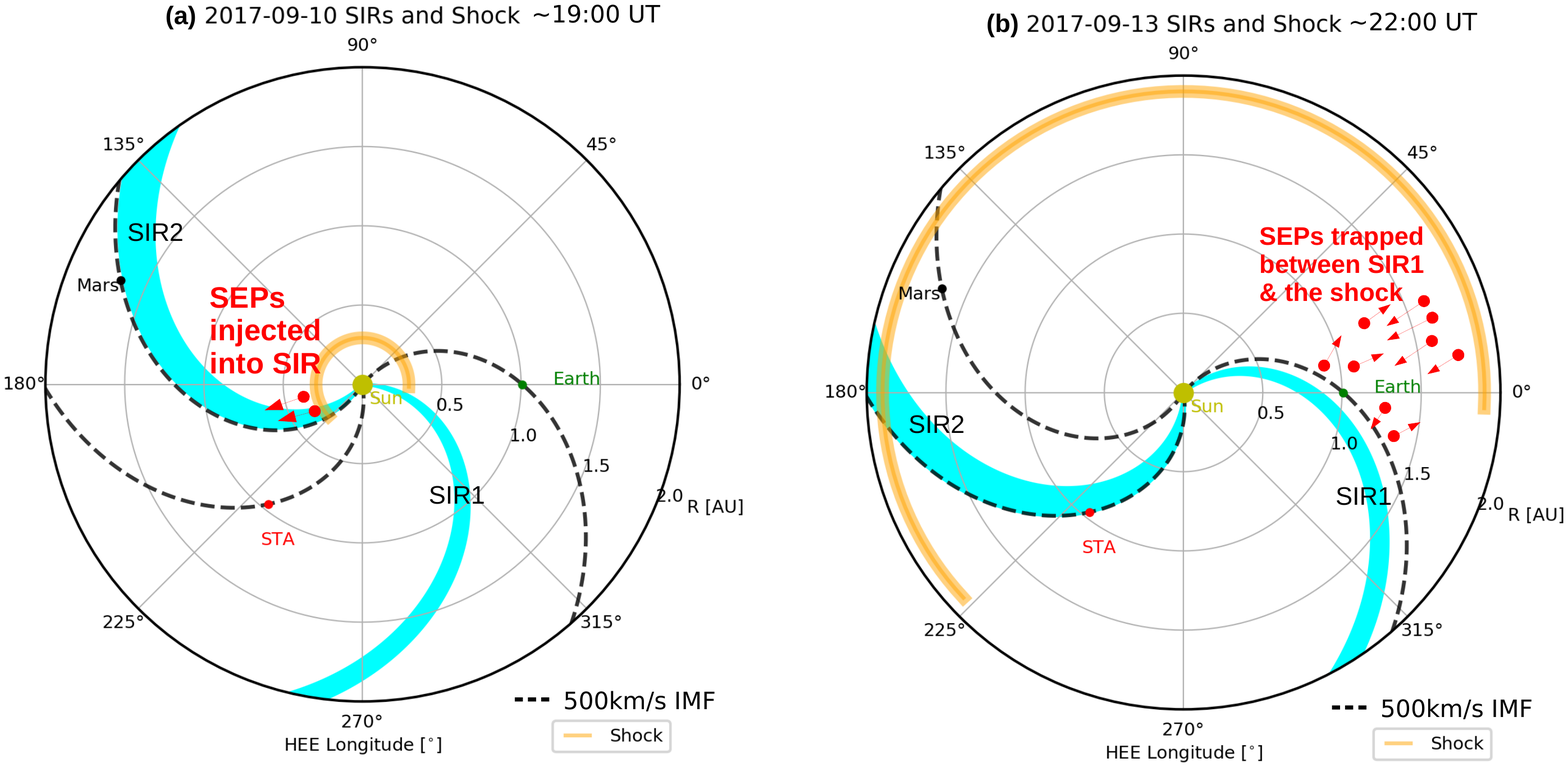}%
	\caption{Illustration of the two SIRs detected at Earth (SIR1) and STA (SIR2) as highlighted in Fig. \ref{fig:insitu}. Both SIRs are time shifted to 2017-09-10 at around {19:30} UT shown in (a) and to 2017-09-13 at around 22:00 UT shown in (b). The widths of SIR1 and SIR2 are derived from {in-situ} their passages at Earth and STA respectively.  Norminal IMFs passing Earth, Mars and STA under a solar wind speed of $\sim$ 500 km/s are plotted in dashed lines. }
	\label{fig:sir}
\end{figure}

\appendix
\section{References of the measurements and databases employed in this study}\label{sec:appendix}
In this appendix, we provide descriptions and references of all the data from various spacecraft, instruments and databases employed in this study.
\begin{enumerate}
	\item {High energetic proton data from the Energetic Proton Electron and Alpha Detector (EPEAD) of the Geostationary Operational Environmental Satellite system 15 (GOES15) have been plotted in Fig. \ref{fig:insitu}(a) and Fig. \ref{fig:dbm}(d). The data are documented at www.ngdc.noaa.gov/stp/satellite/goes/ and are publicly available at https://satdat.ngdc.noaa.gov/.}
	\item {Earth ground-based Neutron Monitors (NMs) measure the secondary particles generated in the atmosphere by primary cosmic energetic charged particles including galactic cosmic rays (GCRs) and solar energetic particles (SEPs). The NM data plotted in Fig. \ref{fig:insitu}(b) are obtained from the Neutron Monitor Data Base (NMDB, www.nmdb.eu/nest/).}
	\item {The Space Physics Data Facility (SPDF) OMNIWeb database (https://omniweb.gsfc.nasa.gov/) provides the solar wind data combined from different measurements of available spacecraft including the Advanced Composition Explorer \citep[ACE,][]{stone1998advanced}, 
		WIND \citep{lin1995wind} and the International Monitoring Platform (IMP)-8. 
		Data plotted in Fig. \ref{fig:insitu}(c)-(e) are in 5 min resolution and the magnetic fields are in the Geocentric Solar Ecliptic (GSE) coordinate.}
	\item {The Radiation Assessment Detector \citep[RAD,][]{hassler2012} measures GCRs and SEPs and their secondaries generated in the Martian atmosphere and regolith at Gale Crater on the surface of Mars since the landing of the Curiosity rover in August 2012. Fig. \ref{fig:insitu}(f) plots the normalized (data divided by the background value) RAD level 2 count rate with the blue curve for particles stopping inside the detector and red curve for particles penetrating through the whole instruments. The average vertical column depth of the atmosphere on top of RAD was about 23.4 g/cm$^2$ during the period of the event. This would only allow protons with kinetic energies larger than about 175 MeV to reach the surface \citep{guo2018generalized} which translates into the minimum primary energy for protons  (SEP on top of the atmosphere) stopping in RAD. For protons to penetrate through the entire detector stack, a minimum energy of 113 MeV is required and this adds to about 288 MeV of primary SEP energy. Note that this approximation is under the assumption that the majority of particles reaching Mars surface are protons which is valid during the SEP events. Fig. \ref{fig:insitu}(g) plots the RAD dose rate [mGray/day] recorded in B (silicon) and E (plastic) detectors. Dose rate is a measure of the energy [$10^{-3}$ Joule] deposited by all detected particles per detector mass [kg] per time unit [day]. A zoomed-in plot (to emphasize the onset of the SEP) of the enhancement ratio of the dose rate (normalized to the background value) in the plastic detector is shown in Fig. \ref{fig:dbm}(c). For downward directed particles during the solar event, they need $\sim$ 100 MeV kinetic energy to reach the E detector. This corresponds to a primary SEP energy $\ge$ about 275 MeV arriving at Mars. Both the count rate and dose rate data are generally in cadence of 17 minutes as RAD runs on an autonomous observing cycle with 16 minutes per observation plus 1 minute of sleep mode.}
	\item {The Analyzer of Space Plasmas and EneRgetic Atoms \citep[ASPERA-3,][]{barabash2006analyzer} experiment of the Mars-EXpress (MEX) mission has been used to derive the solar wind properties \citep{ramstad2015martian} plotted in Fig. \ref{fig:insitu}(h), including the proton density, temperature and solar wind speed.}
	\item {The High Energy Telescope \citep[HET,][]{von2008high} on the Solar Terrestrial relations observatory Ahead (STA) spacecraft provides proton flux rate in various energy ranges which are combined into 4 different channels and plotted in Fig. \ref{fig:insitu}(i). The STA Plasma and Suprathermal Ion Composition \citep[PLASTIC,][]{galvin2008PLASTIC} measurement of the solar wind properties (proton density, temperature and solar wind speed) is shown in Fig. \ref{fig:insitu}(j). The In situ Measurements of Particles And CME Transients (IMPACT) data of the magnetic field experiment \citep{acuna2008stereo} on STA are plotted in Fig. \ref{fig:insitu}(k) in the spacecraft Radial-Tangential-Normal (RTN) coordinate. }
	\item {Remote-sensing coronagraph images of the Sun at two different heliospheric locations have been obtained from a) the Large Angle and Spectrometric Coronagraph \citep[LASCO,][]{brueckner1995large} instrument onboard the Solar and Heliospheric Observatory (SOHO) at Earth and b) the coronagraph (COR) data from the Sun Earth Connection Coronal and Heliospheric Investigation \citep[SECCHI,][]{howard2008} at STA. }
	\item {Observations of extreme-ultraviolet (EUV) phenomena in the solar corona are shown in Fig. \ref{fig:theSun}(a) and (b). They are from the Solar Ultraviolet Imager (SUVI) on the GOES16 spacecraft and the Atmospheric Imaging Assembly \citep[AIA,][]{lemen2011} on board the Solar Dynamics Observatory (SDO) spacecraft.}
\end{enumerate}

\end{document}